

\input eplain

\newcount\fignumber
\def\figdef#1{\global\advance\fignumber by 1 \definexref{#1}{\number\fignumber}{figure}\ref{#1}}
\def\figdefn#1{\global\advance\fignumber by 1 \definexref{#1}{\number\fignumber}{figure}}
\let\figref=\ref
\let\figrefn=\refn
\let\figrefs=\refs

\newcount\tabnumber
\def\tabdef#1{\global\advance\tabnumber by 1 \definexref{#1}{\number\tabnumber}{table}\ref{#1}}
\def\tabdefn#1{\global\advance\tabnumber by 1 \definexref{#1}{\number\tabnumber}{table}}

%
\ifx\pdfoutput\undefined
\input epsf

\def\figscale#1#2{\epsfxsize=#2\epsfbox{#1.eps}}
%
\else

\def\figscale#1#2{\pdfximage width#2 {#1.pdf}\pdfrefximage\pdflastximage}
\fi


\newcount\scount \scount=0



\makeatletter
\def\section#1\par{
  \vskip\z@ plus.3\vsize\penalty-250
  \vskip\z@ plus-.3\vsize\bigskip\vskip\parskip
  \global\advance\scount by1
  \writenumberedtocentry{section}#1{}
  \definexref#1{\the\scount}{section}
  \message{#1}
  \noindent\the\scount.\quad{\bf #1}\nobreak\smallskip\noindent}
\makeatother

\centerline{\bf{Oseen Flow in Paint Marbling}}
\centerline{Aubrey G. Jaffer}
\centerline{agj@alum.mit.edu}

\beginsection{Abstract}

{\narrower

Paint marbling refers to techniques for creating intricate designs in
colored paints floating on a liquid surface.  If the marbling motions
are executed slowly, then this layer of paints can be modeled as a
two-dimensional incompressible Newtonian fluid.

In this highly constrained model many marbling techniques can be
exactly represented by closed form homeomorphisms.  Homeomorphisms can
be composed and compute the composite mapping at any resolution.
Computing homeomorphisms directly is orders of magnitude faster than
finite-element methods in solving paint marbling flows.

Most marbling patterns involve drawing rakes from one side of the tank
to the other; and these can be modeled by exact closed form
homeomorphisms.  But pictorial designs for flowers and animals use
short strokes of a single stylus; presented is an exact velocity field
for Oseen fluid flow and its application to creating short stroke
marbling homeomorphisms.
\par}

\beginsection{Keywords}

{\narrower paint marbling; Oseen flow; Stokes flow; fluid mechanics\par}

\beginsection{Table of Contents}

\readtocfile

\section{Introduction}

Marbling originated in Asia as a decorative art more than 800 years
ago and spread to Europe in the 1500s where it was used for endpapers
and book covers.

The mathematical fascination with paint marbling is that while rakings
across the tank stretch and deform the paint boundaries, they do not
break or change the topology of the surface.  With mechanical guides,
a raking can be undone by reversing the motion of the rake to its
original position.  Raking is thus a physical manifestation of a
homeomorphism, a continuous function between topological spaces (in
this case between a topological space and itself) that has a
continuous inverse function.

\section{Dropping Paint}

First, paints are dropped onto the surface.  Consider the tank as being
an infinite plane covered with a film of ``empty paint'' initially.  The
first paint drop forms a circular spot with area $a$.  If a second drop
with area $b$ is put in the center of the first drop, then the total
covered area increases from $a$ to $a+b$.  Points near the center will
move from small radius to radius $\sqrt{a/\pi}$; and boundary points
will move from radius $\sqrt{a/\pi}$ to radius $\sqrt{(a+b)/\pi}$.

The movements of a point on the surface do not depend on its paint
color; so the motion of every point is the same as for the concentric
paint-drop case.  Given a point $\vec P$ and a new paint drop of radius
$r$ centered at $\vec C$, map the point $\vec P$ to:
$$\vec C+\left(\vec P-\vec C\right)\sqrt{1+{r^2\over \left\|\vec P-\vec C\right\|^2}}$$
\figref{drops} shows a pattern formed by serial injection of 75 drops
of random size and position.  For a more complete discussion of
paint-dropping see {\it Mathematical
Marbling}\cite{10.1109/MCG.2011.51}.

\vbox{\settabs 2\columns
\+\hfill\figscale{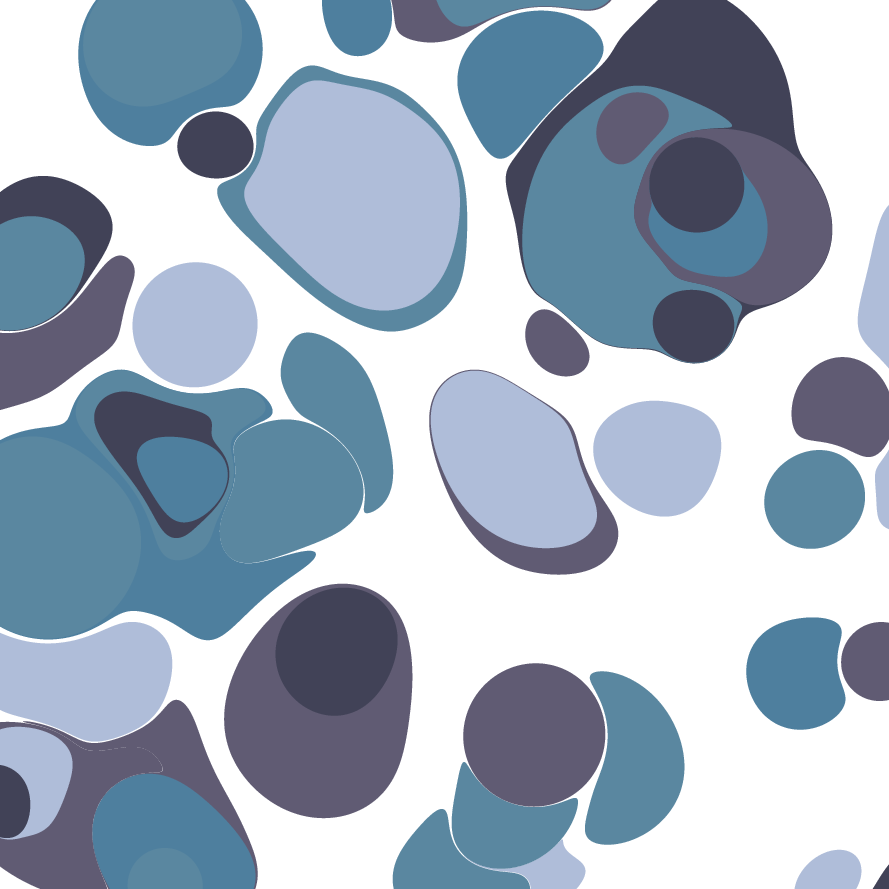}{200pt}\hfill&\hfill\figscale{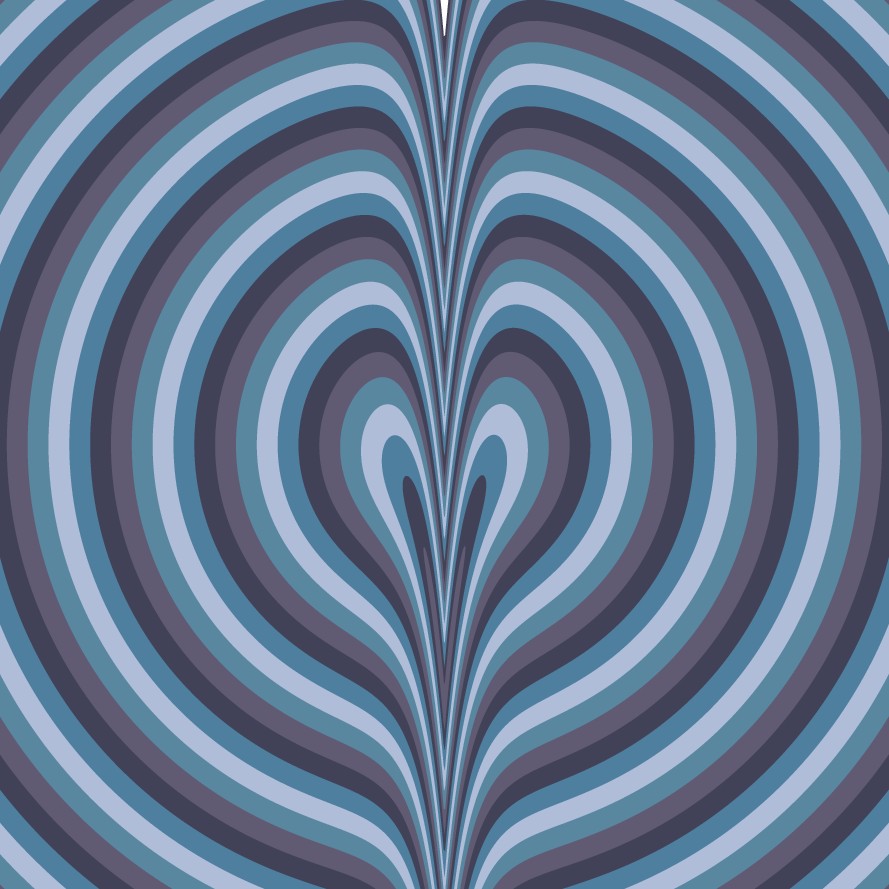}{200pt}\hfill&\cr
\+\hfill\figdef{drops}\hfill&\hfill\figdef{line}\hfill&\cr
}

\section{Line Deformation}

Consider a convex stylus (like a cylinder) partially submerged in the
liquid in the tank.  The system characteristic length $D$ is the
submerged volume of the stylus divided by its wetted surface area.
$V$ is the stylus velocity.  $\nu$ is the kinematic viscosity of the
tank liquid.

Immediately adjacent to the line of stylus motion, parcels of fluid
are moved a distance $V\,t$.  Because flow is laminar, flow must be
uniform along the line.  The friction between adjacent lamina results
in an exponential decrease with (perpendicular) distance from the
line.  For a line on the $y$-axis, the displacement in the
$y$-direction would be:
$$V\,t\exp{-|x|\over L}$$

The characteristic length $L$ is a distance.  While $\nu/V$ is a
distance, its dependence on $\nu$ and $V$ are incorrect, and $L$
should increase with $D$.  Letting $L$ be proportional to
$V\,D^2/\nu={\rm Re}\,D$ is a distance with the correct dependencies.

Because the displacement depends only on distance in the perpendicular
direction, the displacements from parallel lines add linearly, which
speeds computation of (parallel) raking homeomorphisms.  \figref{line}
shows a single line deformation through the center of concentric paint
circles.

A line with unit direction vector $\vec M$ and point $\vec B$ on the
line maps point $\vec P$ to:
$$\vec P+V\,t\,\vec M\exp{\left\|\left(\vec P-\vec B\right)\times\vec M\right\|\over-L}$$




\section{Short Stroke}

At the 2016 Lowell Folk Festival Regina and Dan St. John (Chena River
Marblers) were kind enough to let me perform an experiment on their
equipment.  With a marbling pattern already in the tank, I took a rod,
inserted it into the tank, moved it a short distance, and withdrew it
from the tank.  I could then clearly see the effect of a short stroke
on the paint contours floating in the tank.  The bands perpendicular to
the rod motion were compressed in the direction of motion and spread
perpendicular to the motion to form a gentle curve.  Behind the point
where the rod was withdrawn the contours formed a sharp V leading to
the extraction point.\numberedfootnote{I was so engaged that I
neglected to take a photograph.}

I repeated the experiment, but stopping the motion halfway, then
resuming for the same total distance; the deformation was
indistinguishable from the first stroke.

A vertical rod drawn slowly through a layer of paints floating on the
surface of a liquid, one of the techniques of paint-marbling, can be
treated as the motion of a circular disk in an incompressible
two-dimensional liquid.  Because the movement is slow, viscous forces
dominate inertial forces, so Re$=|U|D/\nu~<~0.1$, where $D$ is the
stylus diameter and $\nu$ is the kinematic viscosity of the liquid.
If paints are floated on a liquid\numberedfootnote{The paint layer is
much thinner than the liquid on which it floats; so the viscosity of
the tank liquid is paramount.} with the viscosity of 10W-40 engine
oil, $\nu\approx10^{-3}{\rm m^2/s}$.  A 2~mm diameter tine moving at
$U=5$~cm/s would result in Re~$\approx0.1$.  A tank of water would
result in Re~$\approx100$.

In the simulation of the two-dimensional
Stokes model\cite{wiki-Stokes} shown in \figref{Stokes-stream}, as the
radius of the cylinder shrinks, the $y$ displacements vanish.  Thus
the Stokes flow cannot produce the spreading I observed ahead of the
motion of the rod.  In the simulation of the Oseen
model\cite{wiki-Oseen} shown in \figref{Oseen-stream} the streamlines
are orbits displacing the $y$ coordinates of points at any distance
from disk.

In ``Boundary-layer theory''\cite{opac-b1124750} Schlichting notes
that, while the Stokes streamlines are symmetrical around the
$y$-axis, the Oseen streamlines are not.  Although subtle, the
asymmetry can be seen in the top and bottom streamlines
in \figref{Oseen-stream} (which is a combination of Stokes and Oseen
flows).  With its symmetry, Stokes flow can be reversed; moving
the cylinder left then right returns the fluid to its initial
position.

\vbox{\settabs 2\columns
\+\hfill\figscale{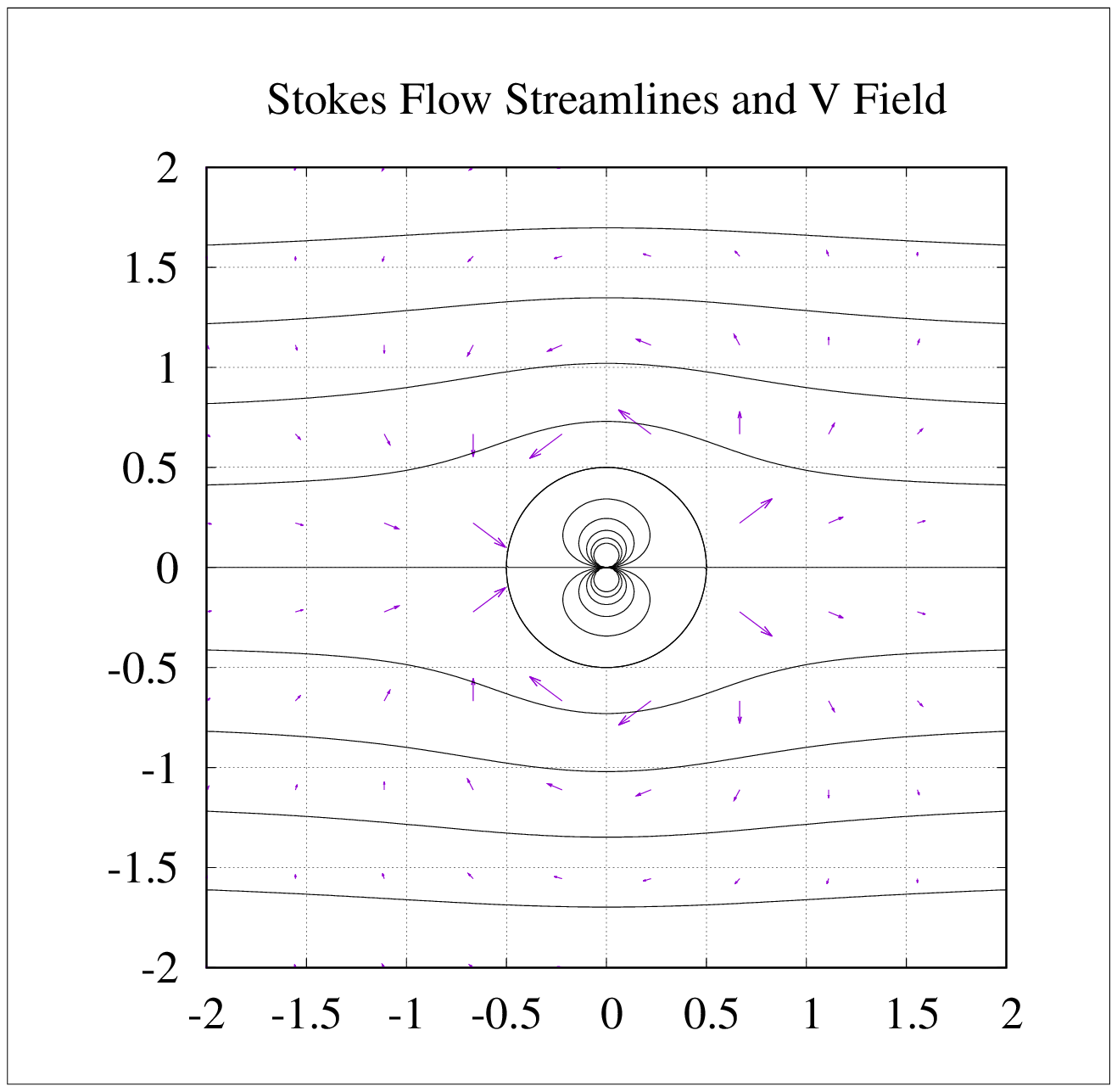}{230pt}\hfill&\hfill\figscale{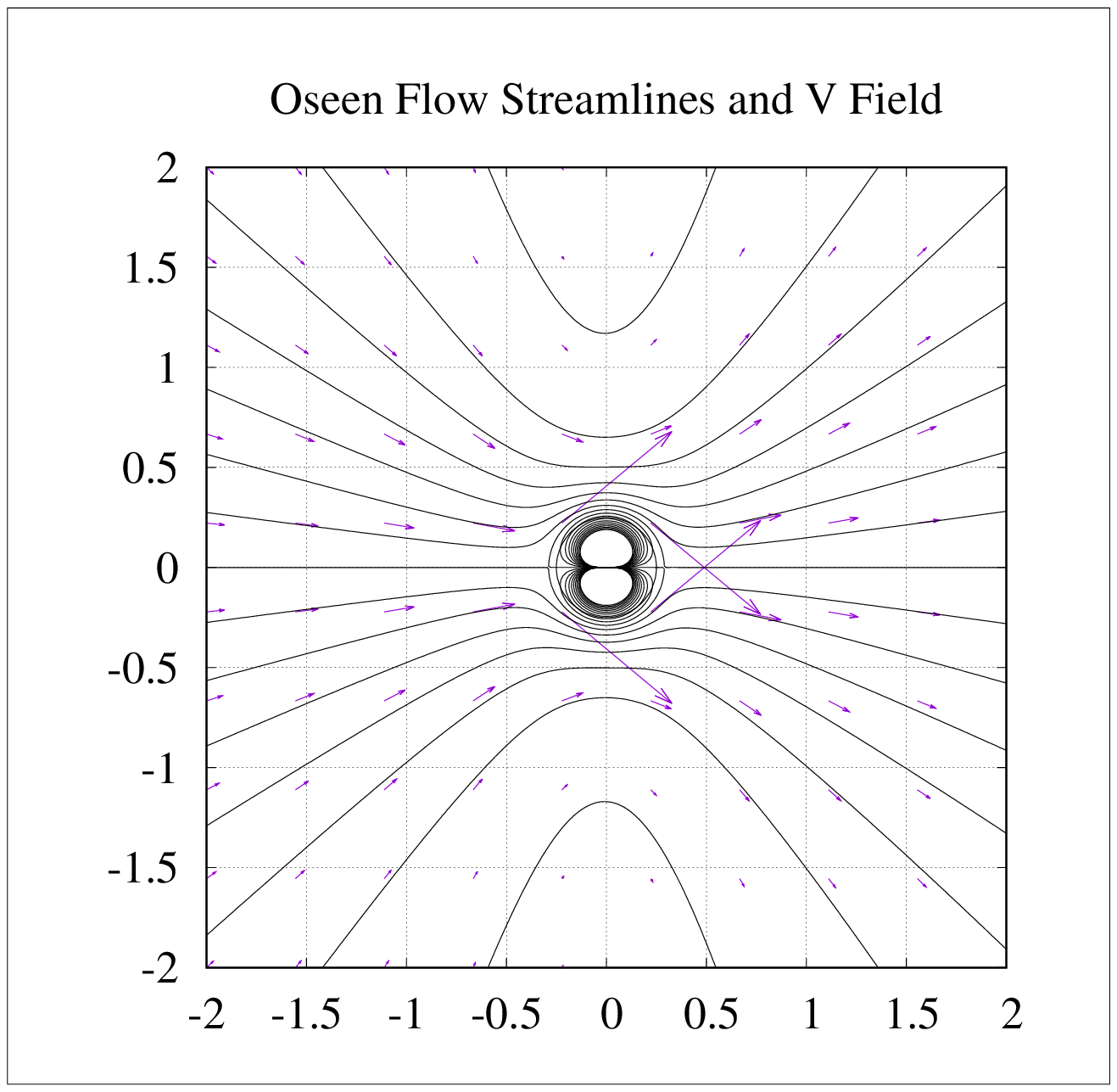}{230pt}\hfill&\cr
\+\hfill\figdef{Stokes-stream}\hfill&\hfill\figdef{Oseen-stream}\hfill&\cr
}

In ``{\it Small $Re$ flows, $\epsilon = Re \ll1$}''
Lagr\'ee\cite{Lagree2015smallRe} gives a derivation of the
two-dimensional Oseen formula which is composed of a near-field and
far-field componenets.  The near field is responsible for the flow
deflecting around the cylinder, while the far field is resonsible for
the wake.  Because the cylinder is withdrawn at the end of the stroke,
the deflection will collapse, and is not of interest for the marbling
deformation.  But the far-field component of the Oseen approximation
results in streamlines with the potential to be more marbling-like.

\section{Velocity Field}

Because the fluid is modeled as incompressible, the divergence of the
velocity field $\nabla\cdot\vec F=0$.  In polar coordinates:

$$\nabla\cdot\vec F(r,\theta)={1\over r}{\partial r F_r\over\partial r}
  +{1\over r}{\partial F_\theta\over\partial \theta}=0\eqdef{incompressible}$$

At large $r$ values, the magnitude of the velocity vanishes.

$$\lim_{r\to\infty} \left\|\vec F(r,\theta)\right\|=0\eqdef{limit}$$

The other boundary condition is the velocity at the origin $\vec
F(0,0)=[U,0]$ where $U$ is the speed.  Converting this constraint to
polar coordinates:

$$\vec F(0,\theta)=F_r(0,\theta)\hat r+F_\theta(0,\theta)\hat\theta
\qquad\hat r=[\cos\theta, \sin\theta]
\qquad\hat\theta=[-\sin\theta, \cos\theta]\eqdef{polar}$$

$$U=F_r(0,\theta)\cos\theta-F_\theta(0,\theta)\sin\theta
\qquad
  0=F_r(0,\theta)\sin\theta+F_\theta(0,\theta)\cos\theta$$

$$ U=F_r(0,\theta)\cos\theta+F_r(0,\theta){\sin^2\theta\over\cos\theta}
 \qquad
 U=-F_\theta(0,\theta){\cos^2\theta\over\sin\theta}-F_\theta(0,\theta)\sin\theta $$


$$ F_r(0,\theta)=U\cos\theta \qquad
   F_\theta(0,\theta)=-U\sin\theta\eqdef{origin} $$

On the basis of the Oseen formula, it is likely that the functions
$F_r(r,\theta)$ and $F_\theta(r,\theta)$ satisfying the constraints
are the product of trigonometric and exponential expressions.  The
argument to the exponential function must be dimensionless.  Although
$-r/D$ is dimensionless, it does not depend on $U$ or $\nu$.
$-r\,U/\nu$ has the wrong dependence on $U$ and $\nu$.
$-r\,\nu\,D^{-2}\,U^{-1}$ is more promising.  Let $L=D^2\,U/\nu$.

A solution \eqref{p-solution} is incompressible \eqref{incompressible}
and satisfies boundary conditions \eqref{limit} and \eqref{origin}.

$$F_r(r,\theta)=U\cos\theta\exp{-r\over L}
 \qquad
  F_\theta(r,\theta)=\left[{r\over L}-1\right]U\sin\theta\exp{-r\over L}\eqdef{p-solution}$$

Expressing \eqref{p-solution} in Cartesian coordinates \eqref{cart0}
according to \eqref{polar} and integrating to find the stream function
$\psi$ \eqref{stream}:

$$r=\sqrt{x^2+y^2}\qquad
  F_x=U{rL-{y^2}\over rL\exp(r/L)}\qquad
  F_y=U{{xy}\over rL\exp(r/L)}\eqdef{cart0}$$
$$\psi(x,y)={Uy\over\exp(r/L)}\eqdef{stream}$$

 \figref{stroke-4} shows a displacement graph resulting from four
evenly spaced applications of the velocity field to a square
grid.\numberedfootnote{The light gray boxes show the horizontal extent
of each stroke; the vertical grid-lines with an initial position to
the right of the beginning of the first stroke are magenta.}  The
deformation matches my description of the short stroke in the
introduction; bands are compressed ahead of the stroke and a vee
trails it.  \figref{stroke-2} shows a displacement graph resulting
from two applications of the velocity field with the same total
displacement.  \figref{stroke-1} shows a displacement graph resulting
from one application of the velocity field to the square grid.
Clearly, the area of (deformed) squares has not been preserved
in \figref{stroke-1}.  \figref{stroke-stream} shows the streamlines
(contours of constant $\psi$ values) and velocity field vectors from
equations \eqref{p-solution} in the steady-state.  The streamlines
form closed orbits.  While \figref{Oseen-stream} was asymmetrical
around the $y$-axis, asymmetry in the streamlines
of \figref{stroke-stream} is not evident.  The streamlines
equation \eqref{stream} is clearly symmetrical around the $y$-axis.

Because the vorticity of $\vec F$ in equation \eqref{vorticity} is
non-zero away from the $x$-axis, the Oseen flow is rotational while
the Stokes flow is not.

$$\nabla\times\vec F={yU\over rL}\left[3-{r\over L}\right]\exp{-r\over L}\eqdef{vorticity}$$


\vbox{\settabs 2\columns
\+\hfill\figscale{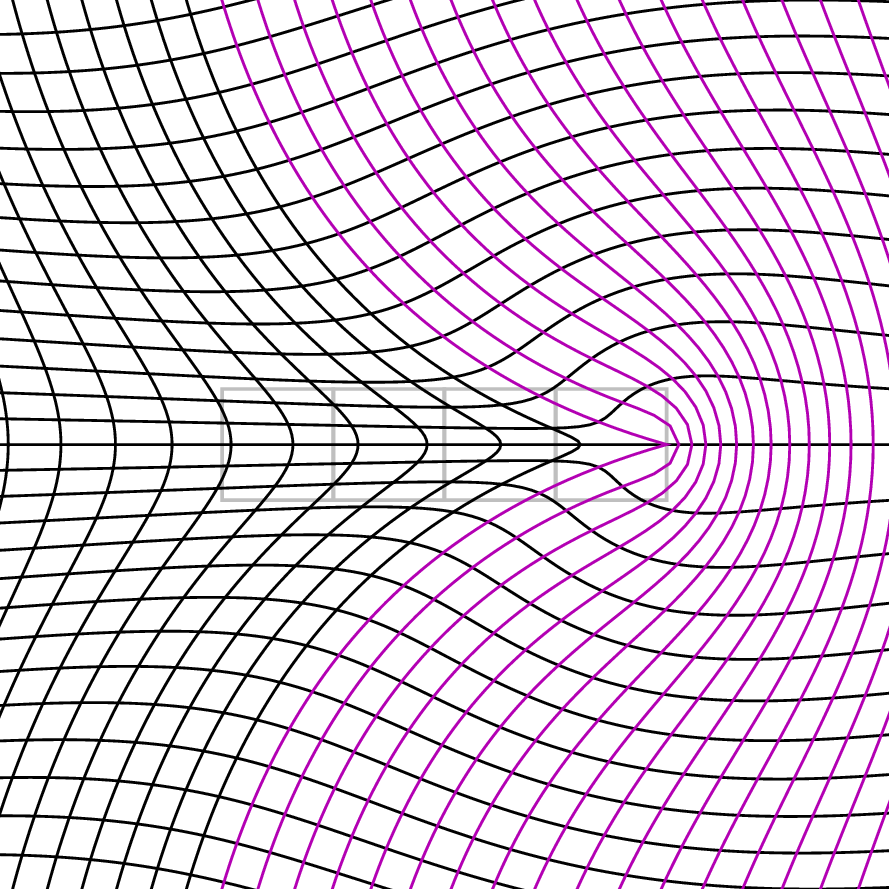}{200pt}\hfill&\hfill\figscale{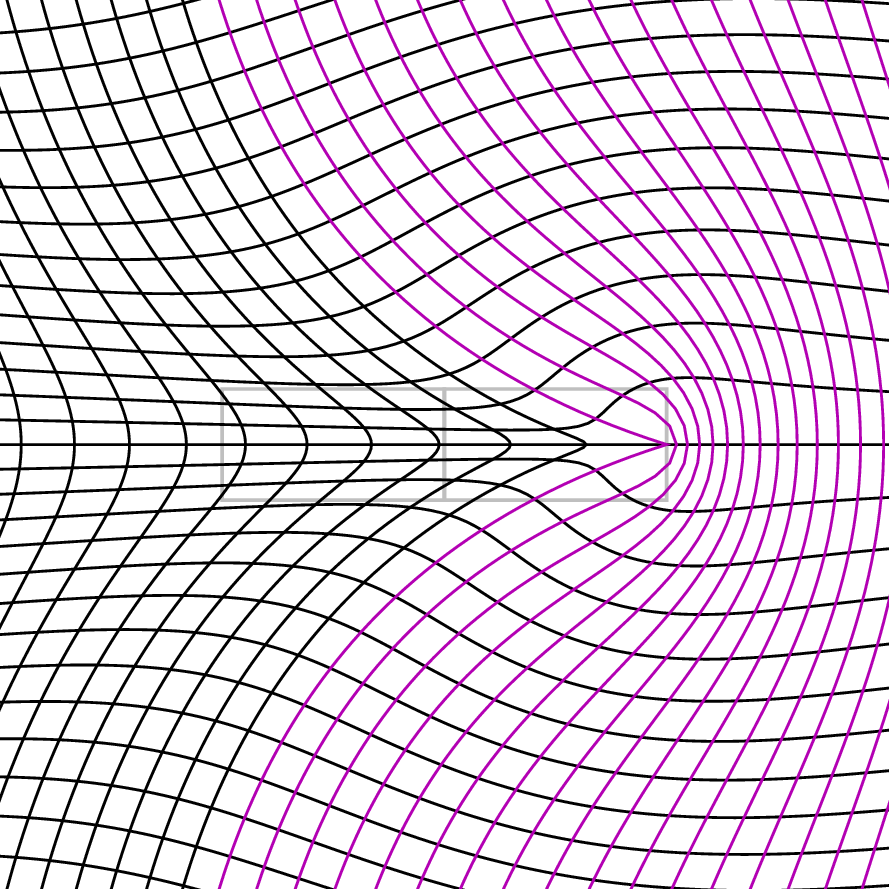}{200pt}\hfill&\cr
\+\hfill\figdef{stroke-4}\hfill&\hfill\figdef{stroke-2}\hfill&\cr
\+\hfill\figscale{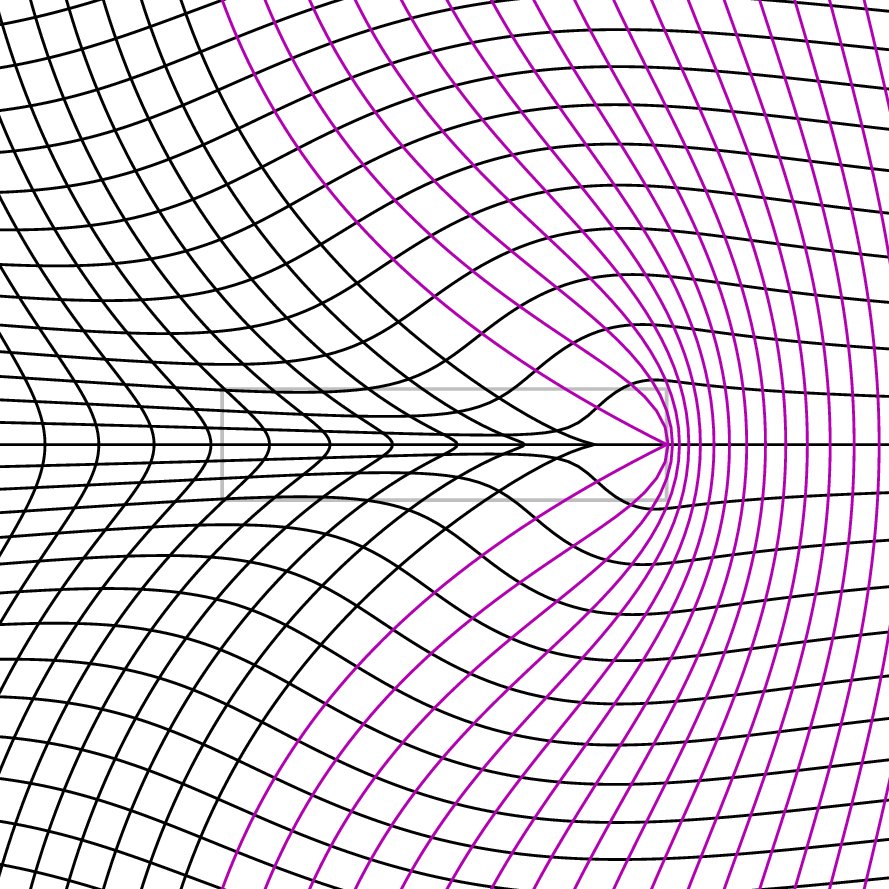}{200pt}\hfill&\hfill\figscale{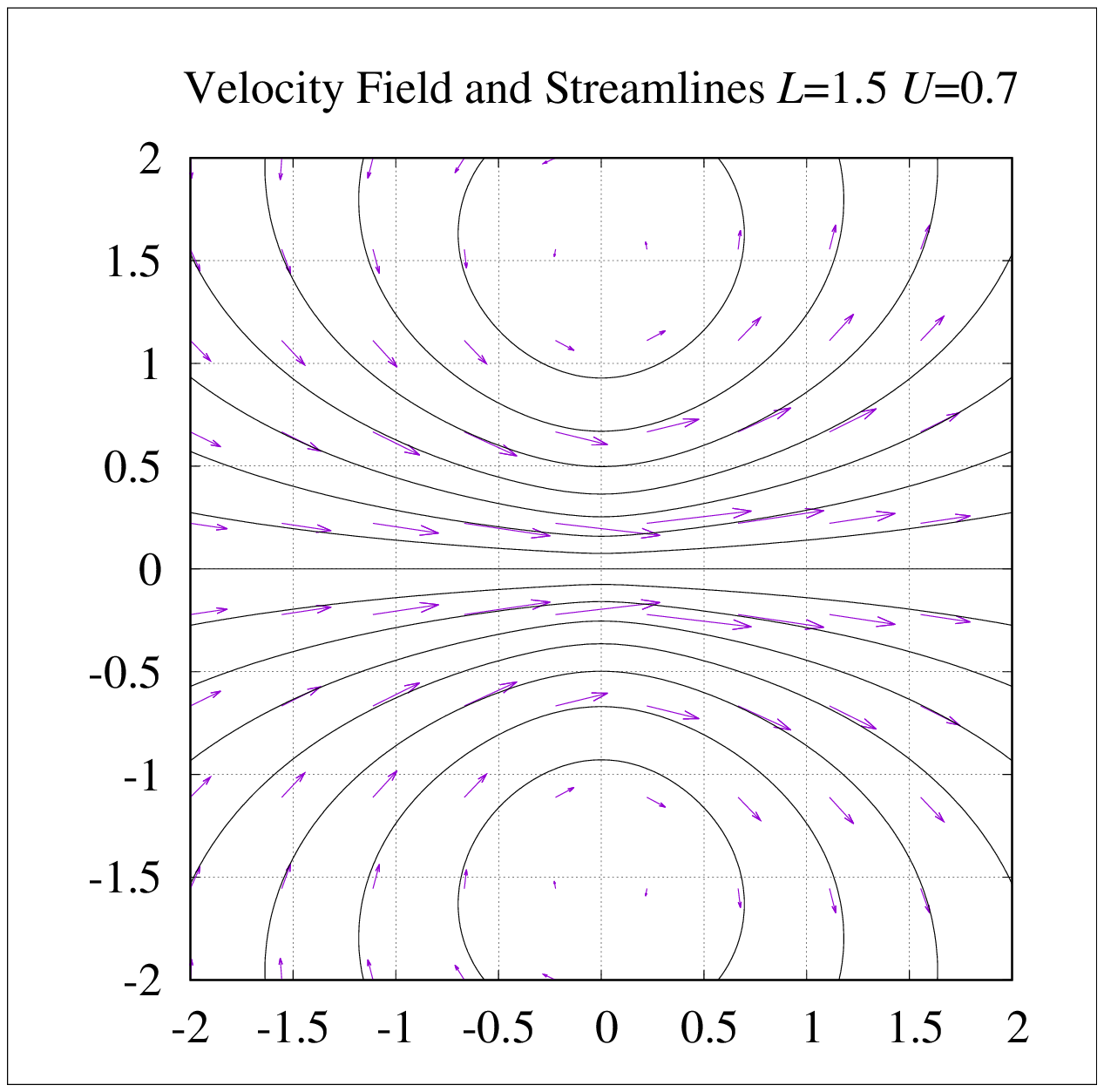}{200pt}\hfill&\cr
\+\hfill\figdef{stroke-1}\hfill&\hfill\figdef{stroke-stream}\hfill&\cr
}


\section{Displacement Field}

The flow velocity field is not the ultimate goal, rather the
displacement after a finite time.  Consider the case of a point on the
horizontal line where $y=0$.  Along this line $x_f>x_0$.  $t(x_f)$ is
time as a function of distance, the inverse of the desired $x_f(t)$.

$$F_x(x,0)=U\exp{-|x|\over L} \qquad F_y(x,0)=0\qquad
  t(x_f)=\int_{x_0}^{x_f}{dx\over F_x(x,0)}$$

If $x_0\ge0$:

$$t(x_f)=\int_{x_0}^{x_f}\exp{|x|\over L}{dx\over U}
  ={L\over U}\left[\exp{x_f\over L}-\exp{x_0\over L}\right]$$

$$x_f(t)=L\ln\left(\exp{x_0\over L}+{t U\over L}\right)
        \eqdef{positive}$$

If $x_0\le x_f\le0$:

$$t(x_f)=\int_{|x_f|}^{|x_0|}\exp{|x|\over L}{dx\over U}
  ={L\over U}\left[\exp{|x_0|\over L}-\exp{-x_f\over L}\right]$$

$$x_f(t)=-L\ln\left(\exp{|x_0|\over L}-{t U\over L}\right)
  =-L\ln\left(\beta\right)\eqdef{negative}$$

Otherwise $x_0\le0$ and $x_f\ge0$:

$$t(x_f)=\int_0^{|x_0|}\exp{|x|\over L}{dx\over U}
        +\int_0^{x_f}\exp{|x|\over L}{dx\over U}
        ={L\over U}\left[\exp{|x_0|\over L}-1
                        +\exp{x_f\over L}-1\right]$$

$$x_f(t)=L\ln\left(2-\exp{|x_0|\over L}+{t U\over L}\right)
  =L\ln\left(2-\beta\right)\eqdef{mixed}$$

$$
  \beta=\exp{|x_0|\over L}-{t U\over L}$$

$x_0\le x_f\le0$ tests the final value of $x_f$ which isn't yet known
when trying to compute it.  But the transition between $x_f(t)$
in \eqref{negative} and $x_f(t)$ in \eqref{mixed} is $\beta=1$; so
test $\beta>1$ instead.

When $x_0\ge0$ formula \eqref{positive} is used.
Formula \eqref{negative} is used when $\beta>1$.
Otherwise formula \eqref{mixed} is used.  \figref{movement} shows the
position versus time of five points on the $x$-axis with $L=1.5$ and
$U=0.7$.

\vbox{\settabs 2\columns
\+\hfill\figscale{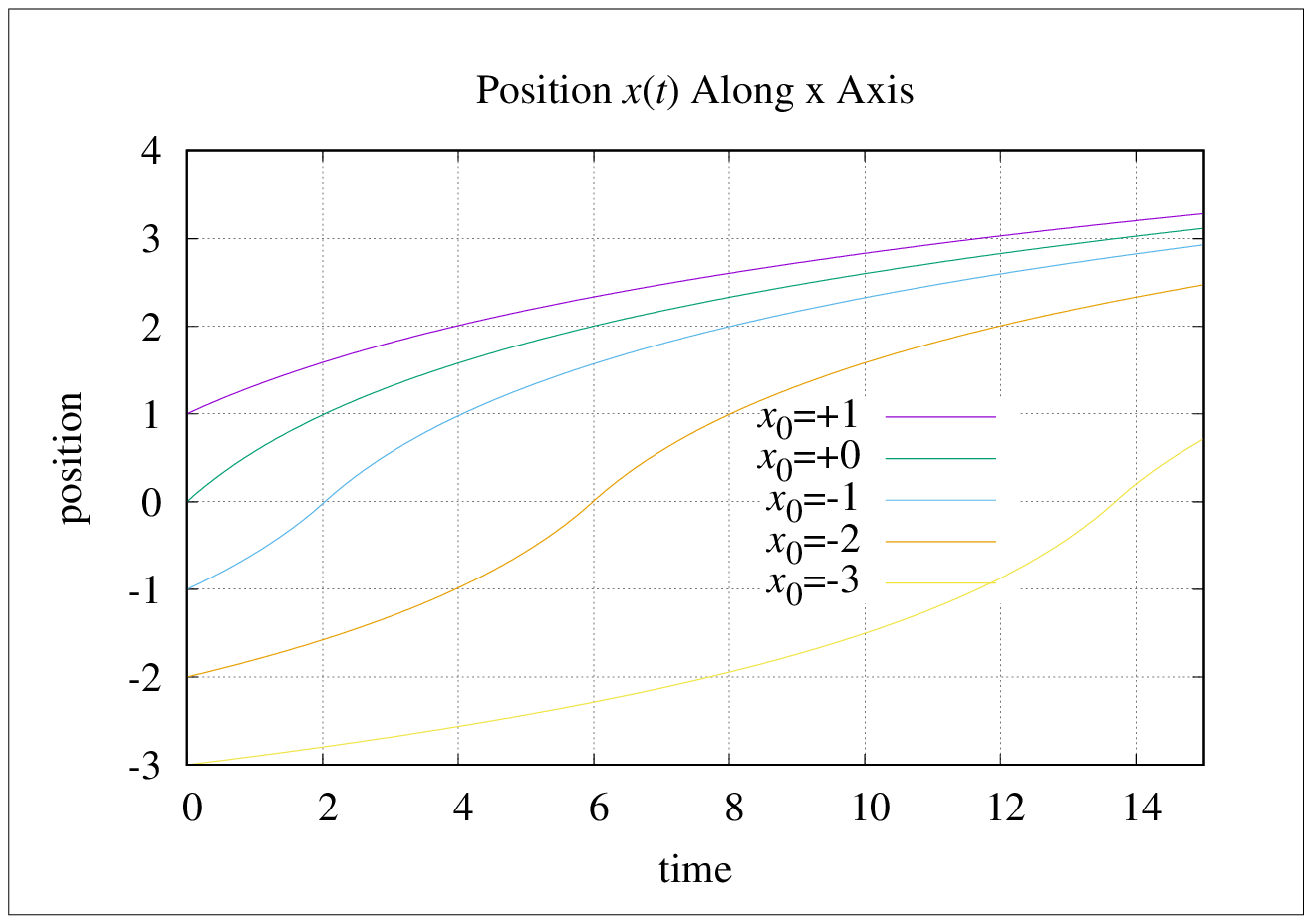}{230pt}\hfill&\hfill\figscale{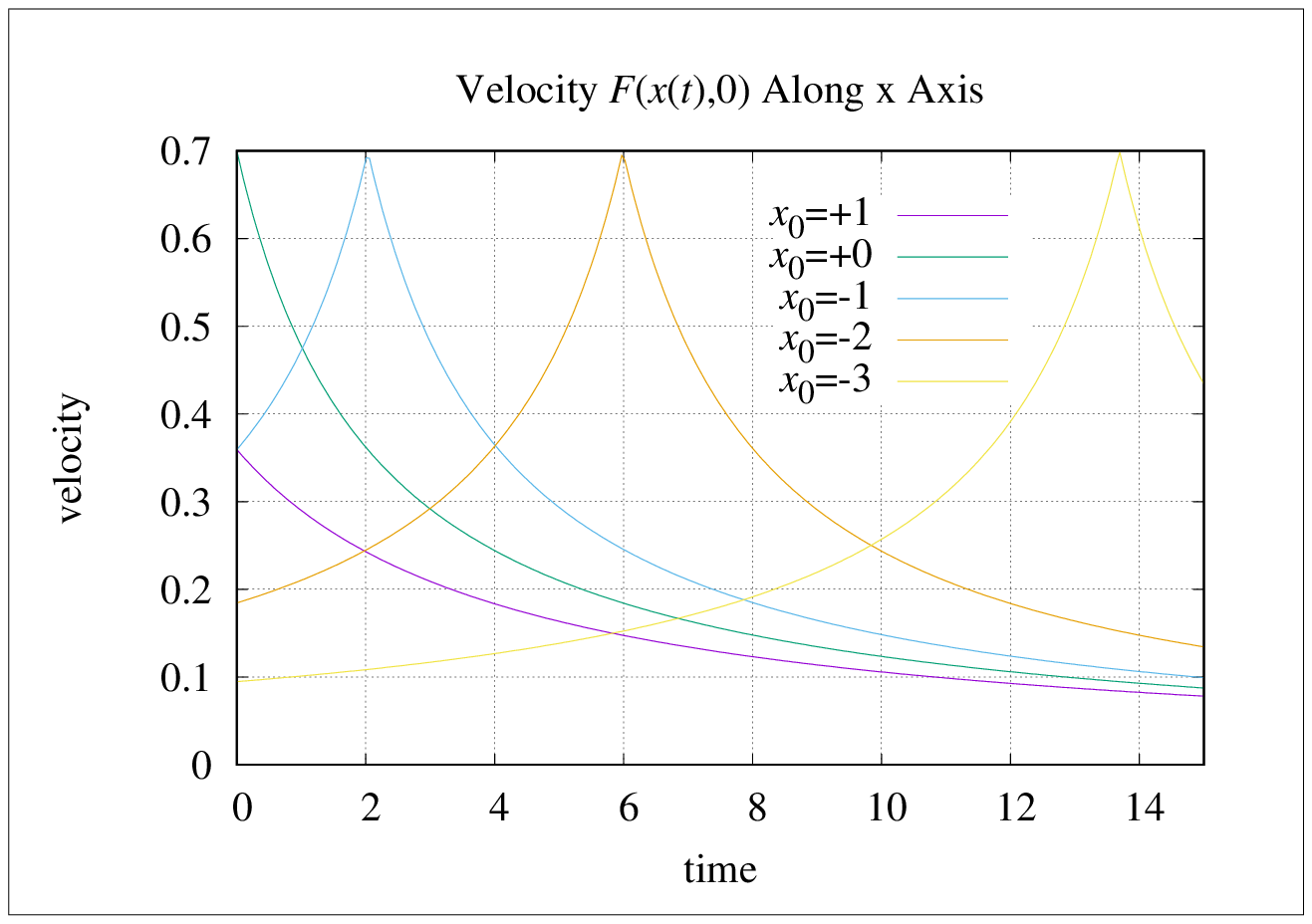}{230pt}\hfill&\cr
\+\hfill\figdef{movement}\hfill&\hfill\figdef{velocity}\hfill&\cr
}

The derivative of $x_f(t)$ with respect to $t$ gives the velocity as a
function of $x_0$ and $t$.  \figref{velocity} shows the velocity
with $L=1.5$ and $U=0.7$.

$${dx_f(t)\over dt}=U\left/
\cases{
{{tU/L}+\exp{|x_0|/L}},&if $x_0>0$;\cr
{-{tU/L}+\exp(|x_0|/L)},&if $-{tU/L}+\exp(|x_0|/L)>1$;\cr
{2+{tU/L}-\exp(|x_0|/L)},&otherwise.\cr
}\right.\eqdef{dxdt}$$

Equation \eqref{dxdt} can be simplified with $r=|x_0|$ and
$\cos\theta=x_0/r$:

$$\eqalign{{dx_f(t)\over dt}&=U\left/
\cases{
{\cos\theta~{tU/L}+\exp({r/L})},&if ${\cos\theta~{tU/L}+\exp({r/L})}>1$;\cr
{2+{tU/L}-\exp(r/L)},&otherwise.\cr
}\right.\cr
&=U\left/\left\{1+\left|\exp{r\over L}+\cos\theta{tU\over L}-1\right|\right\}\right.\cr}$$

This one-dimensional case was (piece-wise) integrable becuase $t(x_f)$
is a monotonic function of a single variable.  \figref{stroke-stream}
shows that all other streamlines are orbits.  Solving for $x$ with a
constant $\psi$ value gives a parameterization of the left ($-$) and
right ($+$) halves of the orbits generated by \eqref{stream}:

$$x_\psi(y)=\pm\sqrt{\left[L\ln{Uy\over\psi}\right]^2-y^2}\qquad y\ne0\eqdef{xrootpsi}$$

The analogous approach for making $\vec F$ and $\psi$ functions of
time is to take the time derivative of the functional inverse of the
integral of the reciprocal of the velocity as a function of $y$
segmented at $x=0$.

Because the orbits and their velocities are continuous, $Uy/\psi$
along the orbits must be either greater or less than
$1$.  \figref{xpsi} shows that $0<Uy/\psi\le1$.

\vbox{\settabs 1\columns
\+\hfill\figscale{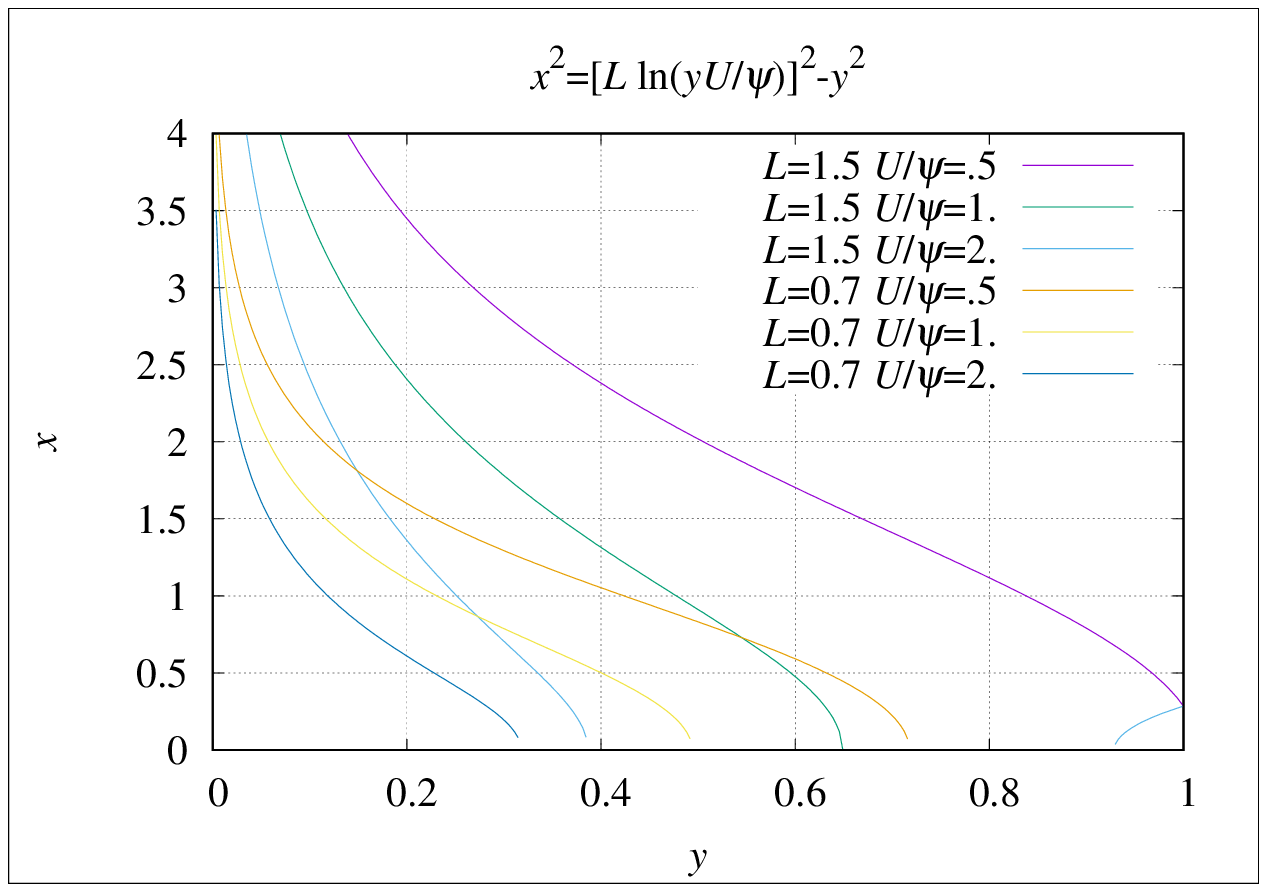}{300pt}\hfill&\cr
\+\hfill\figdef{xpsi}\hfill&\cr
}

Let $w(y)$ be the magnitude of velocity:

$$\eqalign{w(y)^2&={F_x(x_\psi(y),y)^2+F_y(x_\psi(y),y)^2}\cr
  &=U^2
  {[L^2+y^2]/L^2-2y^2\left/\left[L\sqrt{L^2\ln(yU/\psi)^2+y^2}\right]\right.
  \over\exp\left(2\sqrt{L^2\ln(yU/\psi)^2+y^2}/L\right)}\cr
  }$$


$$\zeta=\sqrt{L^2\ln(yU/\psi)^2+y^2}\qquad \int{dy\over w(y)}=\int{L\zeta\exp\left(\zeta/L\right)dy\over U\sqrt{\left[L^2+y^2\right]\zeta^2-2L\zeta y^2}}$$

It seems unlikely that $dy/w(y)$ will be integrable and even less
likely that the displacement field will be expressible in closed form.

\vbox{\settabs 2\columns\+
\hfill\figscale{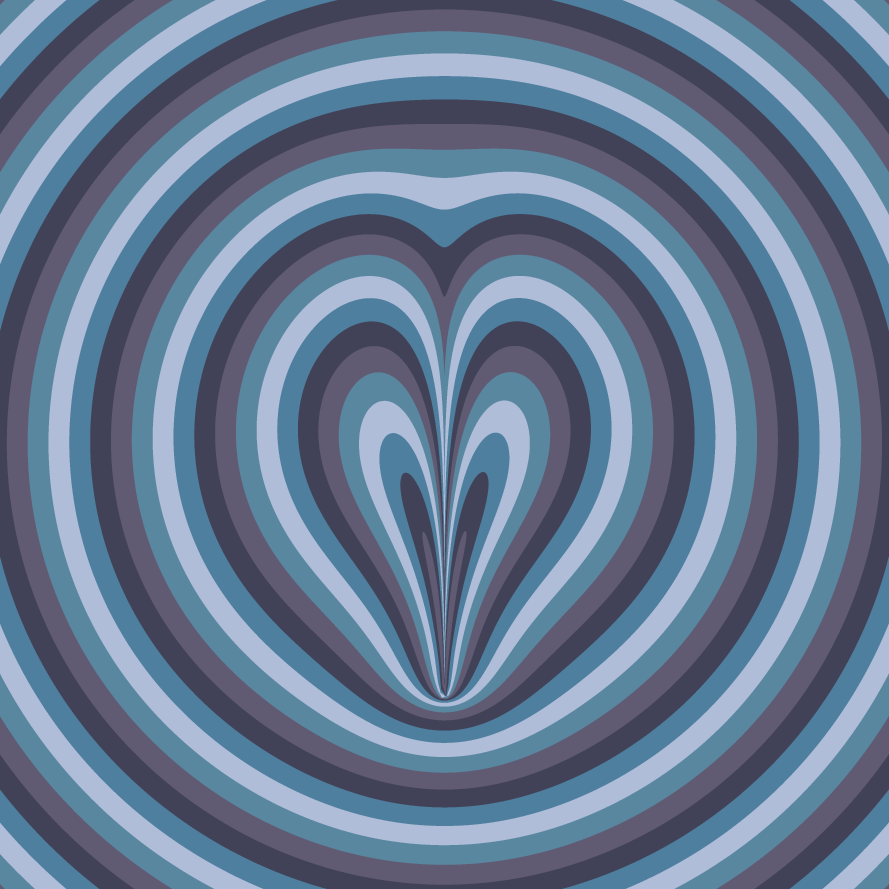}{200pt}\hfill&
\hfill\figscale{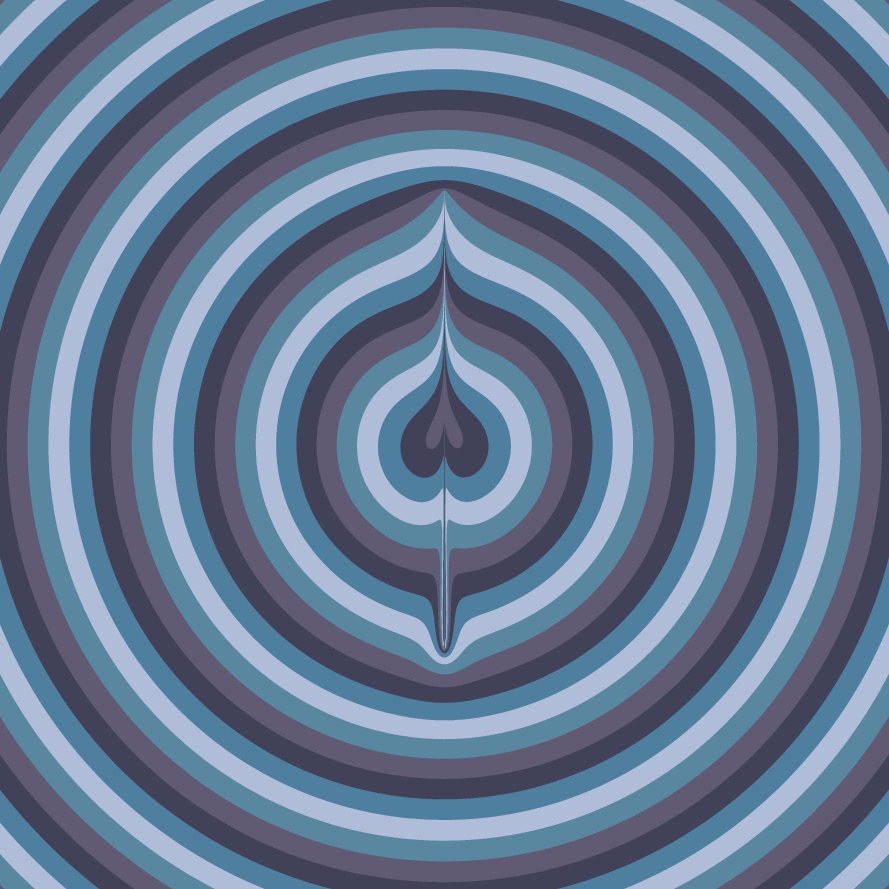}{200pt}\hfill&
\cr\+
\hfill\figdef{strokeink}\hfill&
\hfill\figdef{strokeinkU}\hfill&
\cr\+
\hfill\figscale{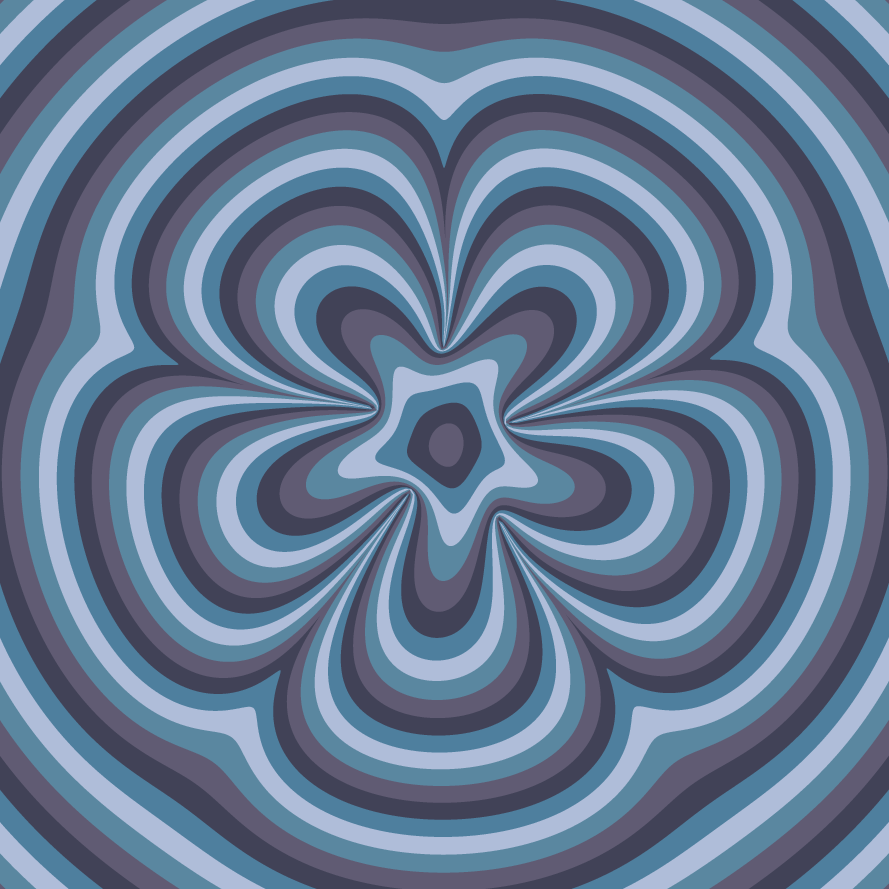}{200pt}\hfill&
\hfill\figscale{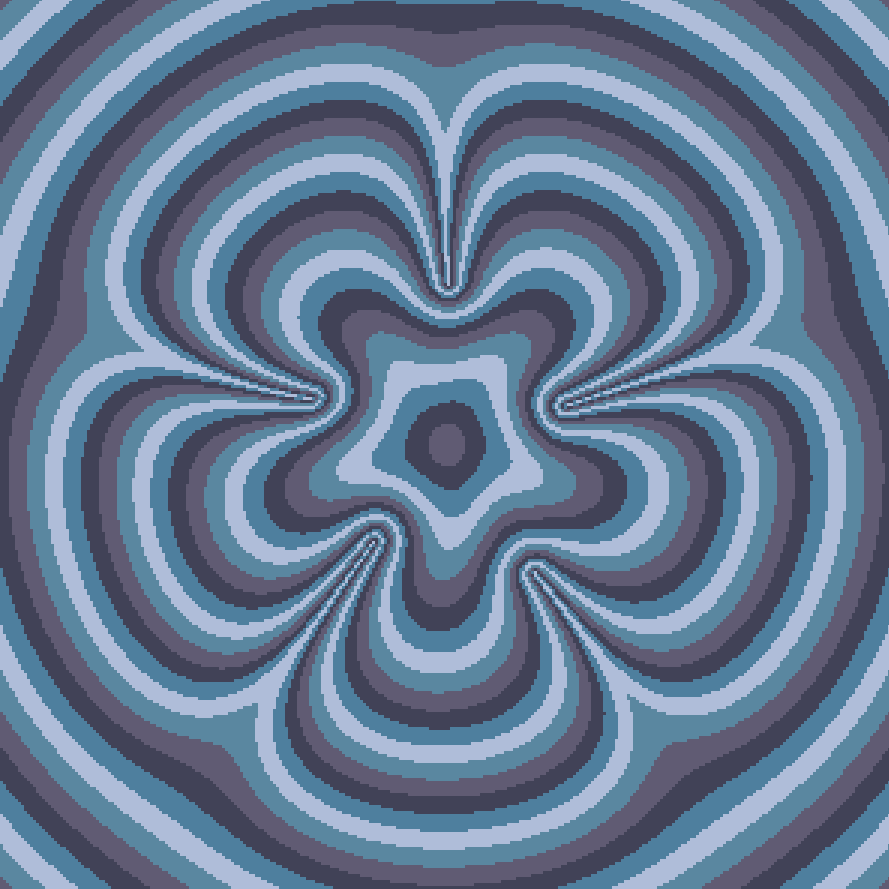}{200pt}\hfill&
\cr\+
\hfill\figdef{strokeink5}\hfill&
\hfill\figdef{strokeink5R}\hfill&\cr
}

\figref{stroke-4} shows that with iteration, the velocity field
generates a reasonable approximation to the displacement field.  It
was found that $\lceil tU/L\rceil=\lceil \lambda/L\rceil$ iterations
produce deformations that are visually acceptable.  \figref{strokeink}
shows a marbling deformation having stroke-length $\lambda=1$ (half
the height of the image) and $L=0.15$.  

With homeomorphisms we can render either by filling contours computed
from the boundaries of paint-drops, or by finding the color mapped from
each point of the display raster (using the inverse homeomorphism).
For the exact homeomorphisms (drops, lines) the geometries rendered
are the same.  But the imperfect reversibility of the stroke
approximation leads to geometric artifacts.
Executing a stroke from $\vec B$ to $\vec E$ followed by a stroke from
$\vec E$ to $\vec B$ does not completely undo the first stroke's
deformation as seen in \figref{strokeinkU}.

\figref{strokeink5} shows a pattern with five radial strokes.  The
pattern is asymmetrical because the strokes were applied in the
sequence 0,2,4,1,3 and each stroke affects the whole space.
\figref{strokeink5R} shows the raster-rendering of the same
five-stroke pattern.  Unlike \figref{strokeink5}, the peaks of the
fourth white band reach nearly all the way in.

\section{Reversibility}

Averaging the velocity field and the negative of the velocity field
with the stroke beginning and end points reversed results in a
velocity field which is more nearly reversible:
$$r=\sqrt{(x-x_B)^2+y^2}\qquad s=\sqrt{(x-x_E)^2+y^2}\qquad tU=x_E-x_B$$
$$\eqalign{
  F_x&={U\over2}\left[\left(1-{y^2\over rL}\right)\exp{-r\over L}+\left(1-{y^2\over sL}\right)\exp{-s\over L}\right]\cr
  F_y&={U\,y\over2\,L}\left[{x-x_B\over r}\exp{-r\over L}+{x-x_E\over s}\exp{-s\over L}\right]
  \cr}\eqdef{carts}
$$

 \figref{stroke-4s} shows a displacement graph resulting from four
evenly spaced applications of the new velocity field to a square grid.
Compared wth \figref{stroke-4}, the point of the vee is rounded and
the tail has slightly more displacement.  A more rounded point is not
inconsistent with the effect of a stylus having finite
width.  \figrefs{stroke-2s} and \figrefn{stroke-1s} show displacement
graphs resulting from two and one applications respectively of the
velocity field with the same total displacement; both are noticeably
distorted.

\vbox{\settabs 2\columns\+
\hfill\figscale{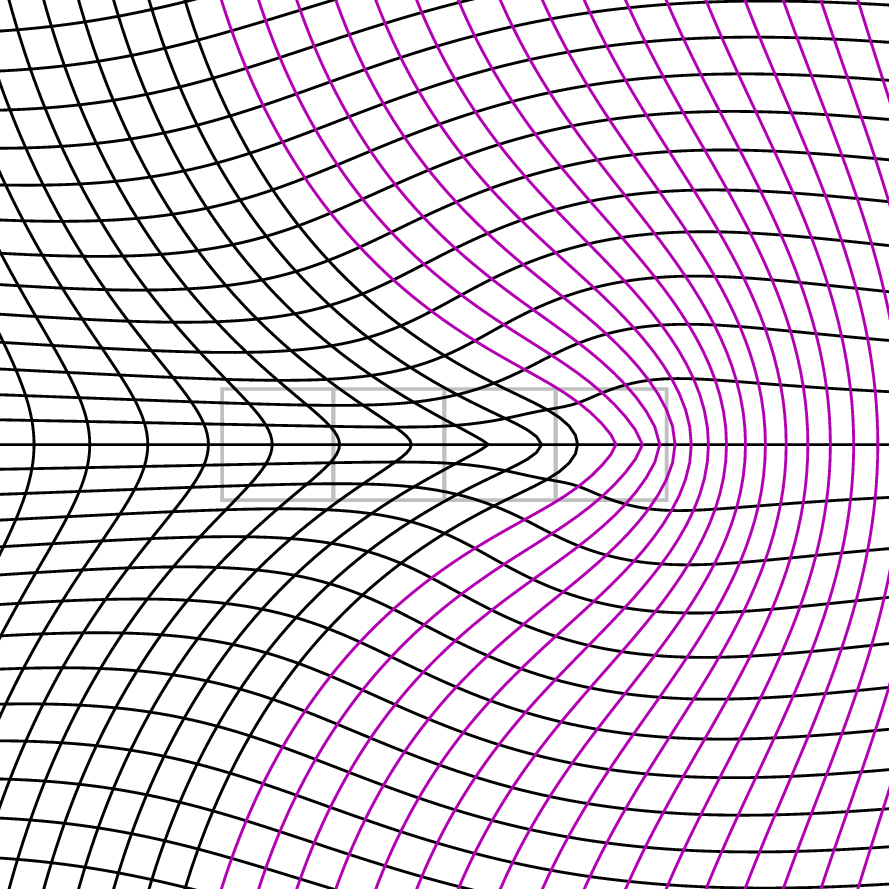}{200pt}\hfill&
\hfill\figscale{stroke-4}{200pt}\hfill&
\cr\+
\hfill\figdef{stroke-4s}\hfill&
\hfill\figref{stroke-4}\hfill&
\cr\+
\hfill\figscale{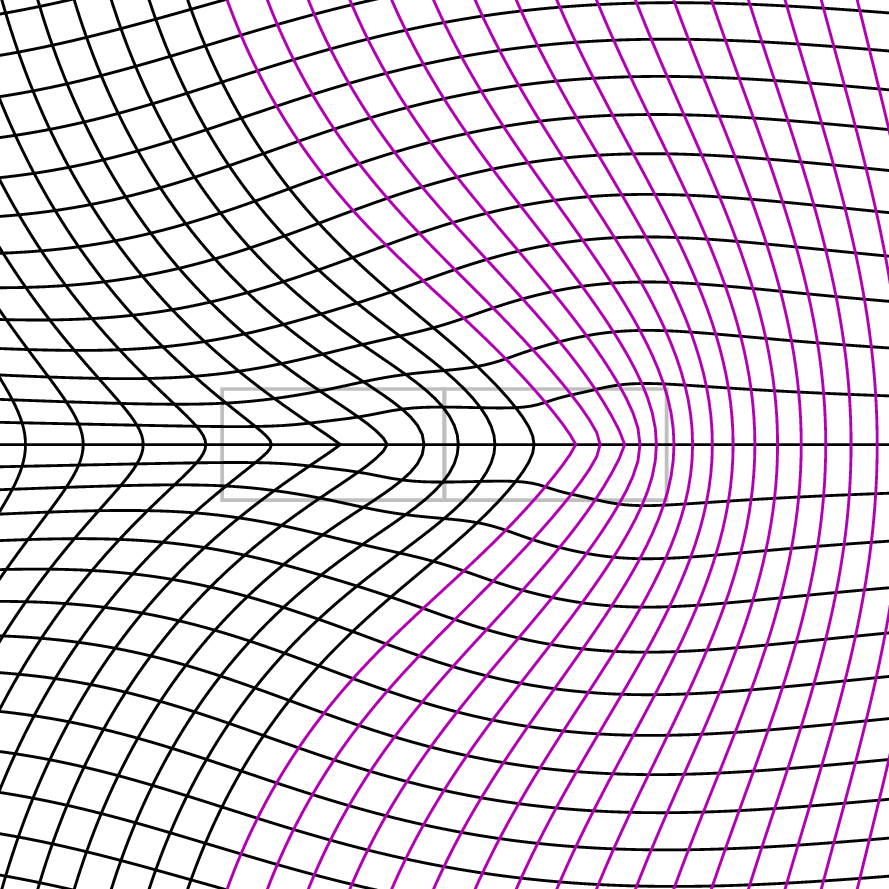}{200pt}\hfill&
\hfill\figscale{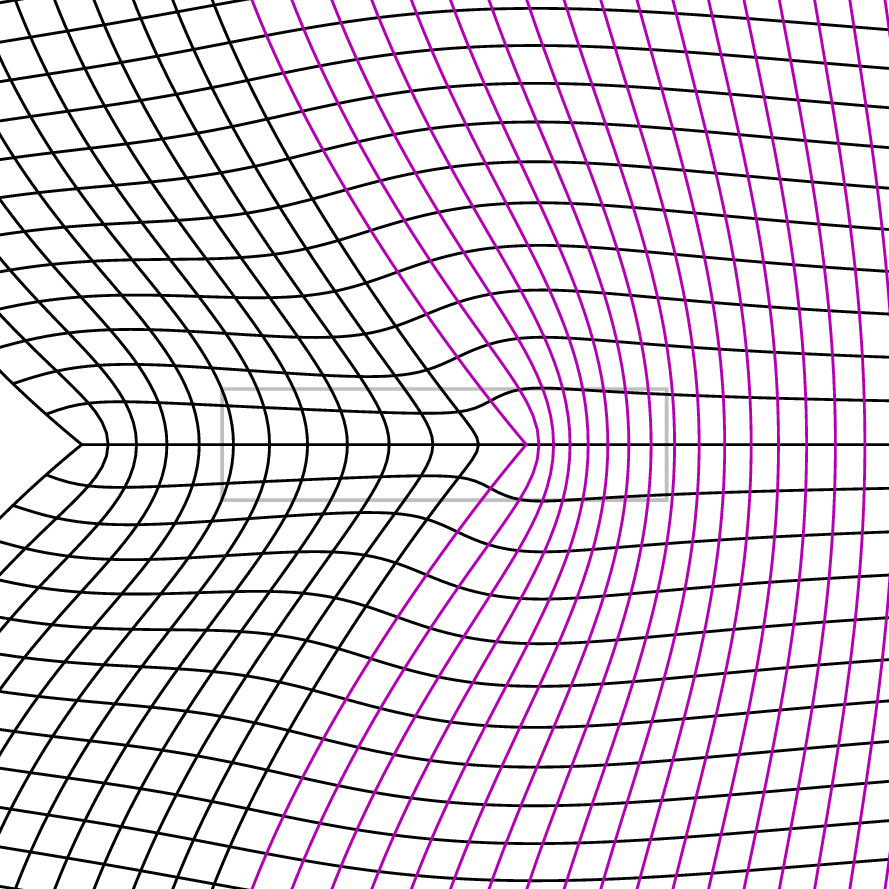}{200pt}\hfill&
\cr\+
\hfill\figdef{stroke-2s}\hfill&
\hfill\figdef{stroke-1s}\hfill&
\cr
}

\section{Application}

This section generalizes velocity field equations \eqref{cart0}
and \eqref{carts} to work at any angle and any stroke length larger
than 0.

Vector function $\vec Q$ maps point $\vec P$ to its new position as a
result of a stroke from $\vec B$ to $\vec E$.  The velocity argument
$U=\lambda/t$.
The system characteristic length $L=V\,D^2/\nu={\rm Re}\,D$ where $D$
is the submerged volume of the stylus divided by its wetted surface
area.

For velocity field \eqref{cart0} the transform is:

$$\eqalign{
\vec Q\left(\vec P,\vec B,\vec E,L\right)&= \vec P+
 \pmatrix{N_x&-N_y\cr N_y&N_x\cr}\cdot\vec F_x(x,y,\lambda/t,L)\cdot t\cr
{\rm where}\qquad 
 \lambda&=\left\|\vec E-\vec B\right\|\cr
\vec N&=\left.\left(\vec E-\vec B\right)\right/\lambda\cr
 x&=\vec N\cdot\left(\vec P-\vec B\right)\cr
 y&=\vec N\times\left(\vec P-\vec B\right)\cr
}$$

For velocity field \eqref{carts} the transform is:

$$\eqalign{
\vec Q\left(\vec P,\vec B,\vec E,L\right)&=\vec P+
 \pmatrix{N_x&-N_y\cr N_y&N_x\cr}\cdot\vec F_x(x_B,x_E,y,\lambda/t,L)\cdot t\cr
{\rm where}\qquad 
 \lambda&=\left\|\vec E-\vec B\right\|\cr
 \vec N&=\left.\left(\vec E-\vec B\right)\right/\lambda\cr
 x_B&=\vec N\cdot\left(\vec P-\vec B\right)\cr
 x_E&=\vec N\cdot\left(\vec P-\vec E\right)\cr
 y&=\vec N\times\left(\vec P-\left.\left(\vec B+\vec E\right)\right/2\right)\cr}$$

The number of segments for the computation is
$n=\left\lceil\left.\left\|\vec E-\vec B\right\|\right/L\right\rceil$ and the
increment vector is $\vec I=\left.\left(\vec E-\vec B\right)\right/n$.

$$\eqalign{
  \vec P&\gets\vec Q\left(\vec P,\vec B,\vec B+\vec I, L\right)\cr
  \vec P&\gets\vec Q\left(\vec P,\vec B+\vec I,\vec B+2\vec I, L\right)\cr
        &\quad\vdots\cr
  \vec P&\gets\vec Q\left(\vec P,\vec B+(n-1)\vec I,\vec B+n\vec I,L\right)\cr
}\eqdef{segmented}$$

 \figref{strokeinks} shows the reversible short stroke; the point is
 not as sharp as \figref{strokeink}.  \figref{strokeinksU} shows the
 reversible short stroke followed by the reverse stroke; cancellation,
 while better than \figref{strokeinkU}, is not complete.

 \figref{strokeinks5} shows the contour-filling
 and \figref{strokeinks5R} shows the raster-rendering versions of the
 five radial stroke pattern.  The difference between them is less than
 between \figrefs{strokeink5} and \figrefn{strokeink5R}.

This segmented computation could be adapted to allow modeling of
curved strokes.

\vbox{\settabs 2\columns\+
\hfill\figscale{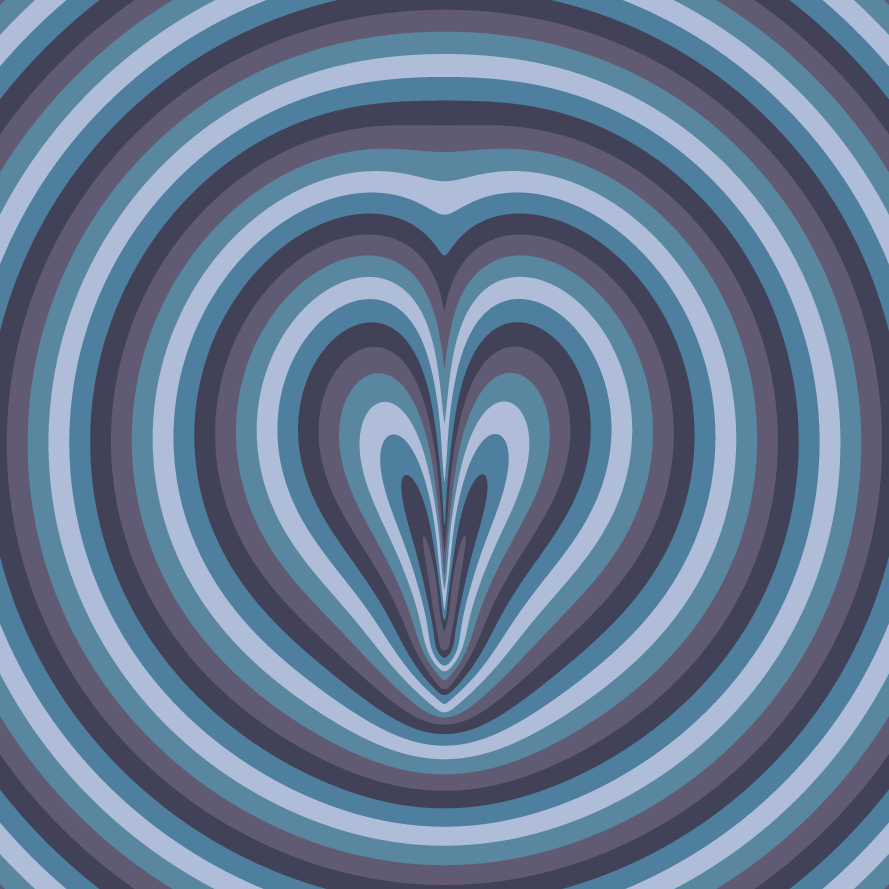}{200pt}\hfill&
\hfill\figscale{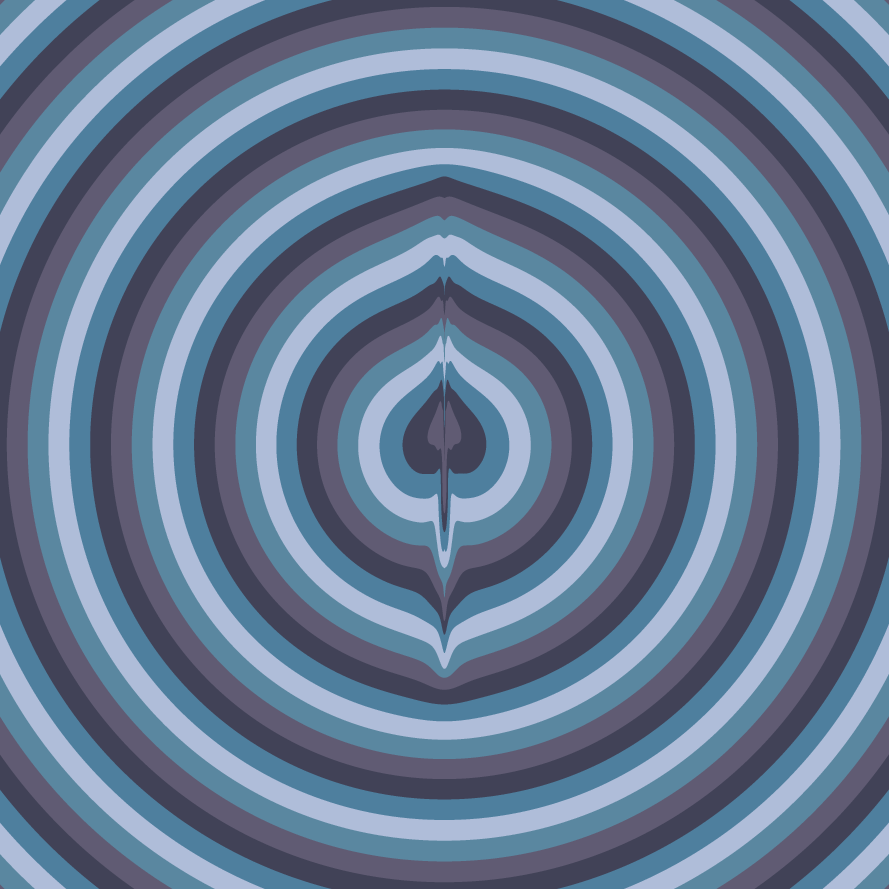}{200pt}\hfill&
\cr\+
\hfill\figdef{strokeinks}\hfill&
\hfill\figdef{strokeinksU}\hfill&
\cr\+
\hfill\figscale{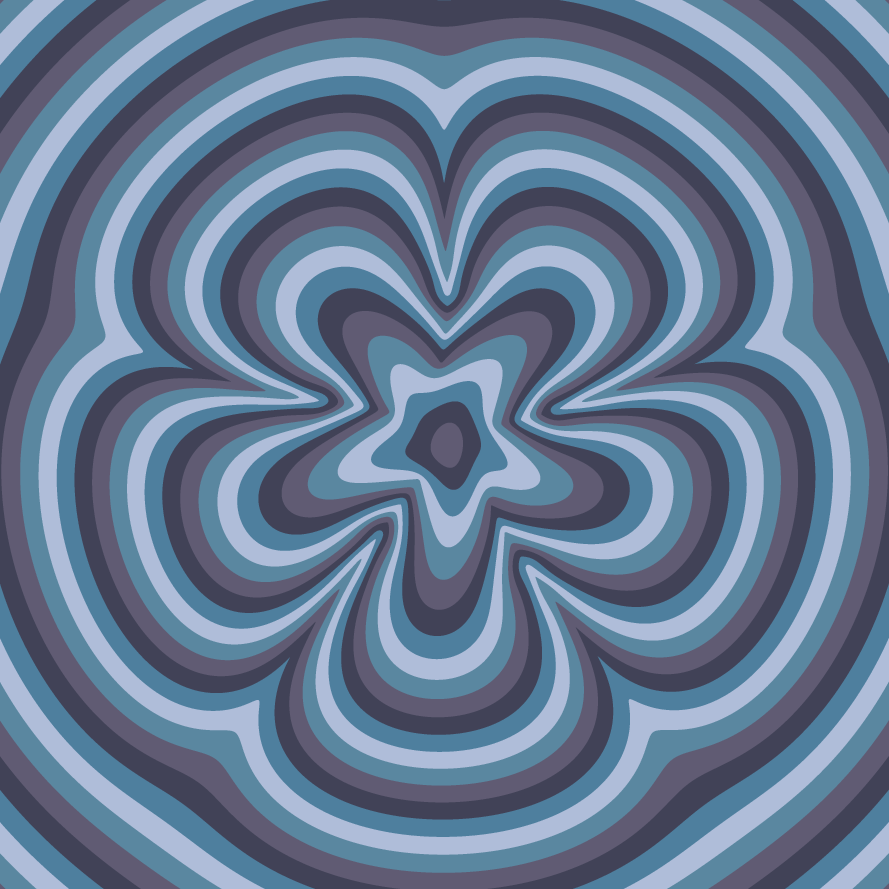}{200pt}\hfill&
\hfill\figscale{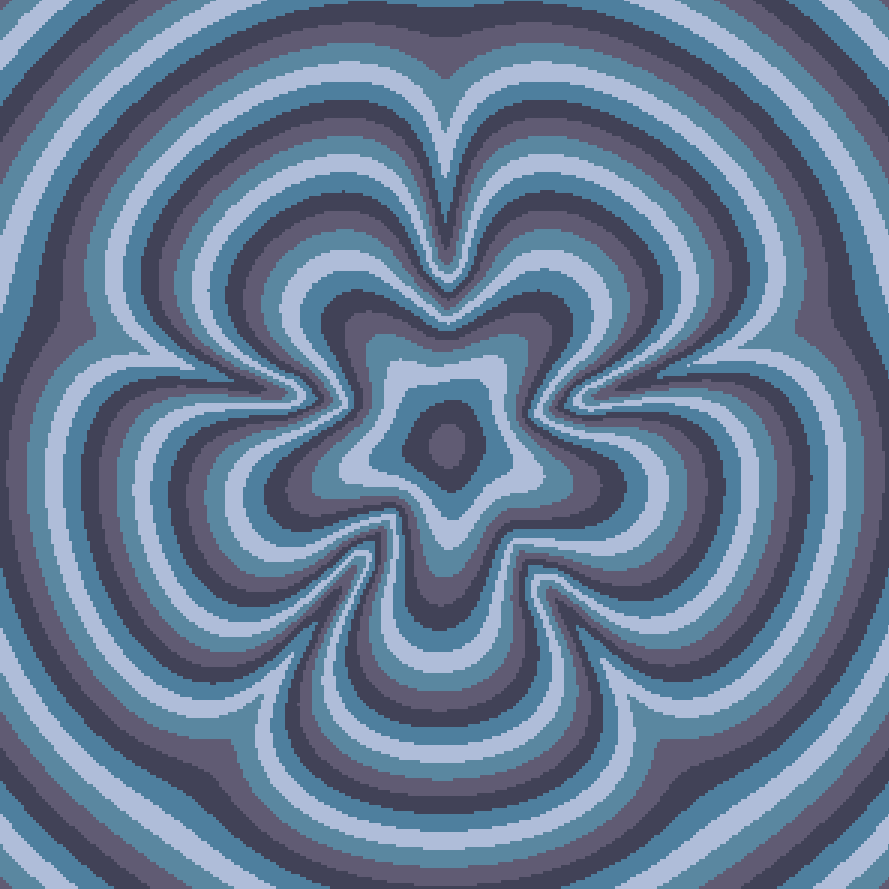}{200pt}\hfill&
\cr\+
\hfill\figdef{strokeinks5}\hfill&
\hfill\figdef{strokeinks5R}\hfill&
\cr
}

\section{Solid Marbling}

{\it Solid Mathematical Marbling}\cite{10.1109/MCG.2016.42} and
{\it Marbling-based Creative Modelling}\cite{Lu:2017:MCM:3106952.3106954}
introduce three-dimensional generalizations of the planar marbling
primitives.  How can these be adapted to finite movements?  An
(unbounded) line can be moved perpendicular to the line.  Because
there will be no flow parallel to the line, the two-dimensional
calculation \eqref{segmented} can be used where the coordinates along
the axis parallel to the line are passed through unchanged.  Similar
reasoning applies to a circle moved along its axis.  Movement of a
point is not the same as the planar case becaue there is rotational
symmetry around the direction of motion; hence the 0 divergence
constraint is different.

In spherical coordinates the direction of flow at the origin is along
the $z$-axis.  The angle $\varphi$ around the $z$-axis does not change
because there is no flow around the $z$-axis.  With the fluid modeled
as incompressible, the divergence of the velocity field
$\nabla\cdot\vec F=0$.  In spherical coordinates\cite{wiki-spherical}:

$$\nabla\cdot\vec F(r,\theta,\varphi)
  ={1\over r^2}{\partial r^2 F_r\over\partial r}
  +{1\over r\sin\theta}{\partial \sin\theta F_\theta\over\partial \theta}
  +{1\over r\sin\theta}{\partial F_\varphi\over\partial \varphi}
  =0 \eqdef{incompressible-sphere}$$

At large $r$ values, the magnitude of the velocity vanishes.

$$\lim_{r\to\infty} \left\|\vec F(r,\theta,\varphi)\right\|=0\eqdef{limit-sphere}$$

The other boundary condition is the velocity at the origin $\vec
F(0,0,0)=[0,0,U]$ where $U$ is the speed.

$$ F_r(0,\theta,\varphi)=U\cos\theta \qquad
   F_\theta(0,\theta,\varphi)=-U\sin\theta \qquad
   F_\varphi(0,\theta,\varphi)=0
   \eqdef{origin-sphere} $$

A solution \eqref{s-solution} very similar to \eqref{p-solution} is
incompressible \eqref{incompressible-sphere} and satisfies boundary
conditions \eqref{limit-sphere} and \eqref{origin-sphere}.
Equations \eqref{sphere} shows the solution in Cartesian coordinates
where $r=\sqrt{z^2+y^2+x^2}$.

$$F_r(r,\theta,\varphi)=U\cos\theta\exp{-r\over L}
 \quad
  F_\theta(r,\theta,\varphi)=\left[{r\over 2L}-1\right]U\sin\theta\exp{-r\over L}\eqdef{s-solution}$$

$$\vec F(r,\theta,\varphi)=F_r(r,\theta,\varphi)\hat r+F_\theta(r,\theta,\varphi)\hat\theta+F_\varphi(r,\theta,\varphi)\hat\varphi$$
$$\hat r=[\sin\theta\cos\varphi, \sin\theta\sin\varphi, \cos\theta]
\quad\hat \theta=[\cos\theta\cos\varphi, \cos\theta\sin\varphi, -\sin\theta]
\quad\hat\varphi=[-\sin\varphi, \cos\varphi, 0]
$$

$$
F_x={xzU\over2Lr}\exp{-r\over L} \qquad
F_y={yzU\over2Lr}\exp{-r\over L} \qquad
F_z=\left[2Lr-y^2-x^2\right]{U\over2Lr}\exp{-r\over L}
\eqdef{sphere}$$

The system characteristic length $L=V\,D^2/\nu={\rm Re}\,D$.  The
entire moving (convex) object being wetted, $D$ is the volume of the
object divided by its surface area.

\beginsection{Acknowledgments}

Thanks to Blake Jones for correcting several problems in the statement
of the algorithms.

\beginsection{References}

\bibliographystyle{unsrt}
\bibliography{stroke}

\vfill\eject
\bye